\documentclass[lettersize,journal]{IEEEtran}
\usepackage{amsmath,amsfonts}
\usepackage{algorithmic}
\usepackage{algorithm}
\usepackage{array}
\usepackage[caption=false,font=normalsize,labelfont=sf,textfont=sf]{subfig}
\usepackage{textcomp}
\usepackage{stfloats}
\usepackage{url}
\usepackage{verbatim}
\usepackage{graphicx}
\usepackage{cite}
\usepackage[colorlinks, linkcolor=red, anchorcolor=red, citecolor=blue]{hyperref}
\usepackage{cleveref}

\usepackage{amsfonts}

\usepackage{epsfig}
\usepackage{amssymb}
\usepackage{pifont}      
\usepackage[accsupp]{axessibility}

\graphicspath{{figures/}}

\usepackage{newfloat}
\usepackage{listings}
\usepackage{multirow}
\usepackage{multicol}
\usepackage{booktabs}
\usepackage{xcolor}

\usepackage{colortbl}%
\newcommand{\myrowcolour}{\rowcolor[gray]{0.925}}

\begin{document}

\title{Dual Degradation Representation for Joint Deraining and Low-Light Enhancement in the Dark}

\author{{Xin Lin*, Jingtong Yue*, Sixian Ding, Chao Ren, Lu Qi and Ming-Hsuan Yang\IEEEmembership{,~Fellow,~IEEE}}

\thanks{*: equal contribution}

\thanks{This work was supported in part by the
National Natural Science Foundation of China under Grant 62171304 and partly by the Natural Science Foundation of Sichuan Province under Grant 2024NSFSC1423, and Cooperation Science and Technology Project of Sichuan University and Dazhou City under Grant 2022CDDZ-09. (Corresponding
author: Chao Ren.)}

\thanks{Xin Lin, Jingtong Yue, Sixian Ding and Chao Ren are with the College of Electronics and Information Engineering, Sichuan University, Chengdu 610065, China (e-mail:
linxin020826@gmail.com; yuejingtong@stu.scu.edu.cn; dingsixian@stu.scu.edu.cn; chaoren@scu.edu.cn).}

\thanks{Lu Qi and Ming-Hsuan Yang are with the University of California at Merced, Merced, CA 95343 USA (e-mail: qqlu1992@gmail.com;
mhyang@ucmerced.edu).}}

\markboth{Journal of \LaTeX\ Class Files,~Vol.~14, No.~8, August~2021}%
{Shell \MakeLowercase{\textit{et al.}}: A Sample Article Using IEEEtran.cls for IEEE Journals}

\maketitle

\begin{abstract}

Rain in the dark poses a significant challenge to deploying real-world applications such as autonomous driving, surveillance systems, and night photography. Existing low-light enhancement or deraining methods struggle to brighten low-light conditions and remove rain simultaneously. Additionally, cascade approaches like ``deraining followed by low-light enhancement'' or the reverse often result in problematic rain patterns or overly blurred and overexposed images. To address these challenges, we introduce an end-to-end model called L$^{2}$RIRNet, designed to manage both low-light enhancement and deraining in real-world settings. Our model features two main components: a Dual Degradation Representation Network (DDR-Net) and a Restoration Network. The DDR-Net independently learns degradation representations for luminance effects in dark areas and rain patterns in light areas, employing dual degradation loss to guide the training process. The Restoration Network restores the degraded image using a Fourier Detail Guidance (FDG) module, which leverages near-rainless detailed images, focusing on texture details in frequency and spatial domains to inform the restoration process. Furthermore, we contribute a dataset containing both synthetic and real-world low-light-rainy images. Extensive experiments demonstrate that our L$^{2}$RIRNet performs favorably against existing methods in both synthetic and complex real-world scenarios. All the code and dataset can be found in \url{https://github.com/linxin0/Low_light_rainy}.

\end{abstract}


\begin{figure}[ht]
  \includegraphics[width=0.47\textwidth]{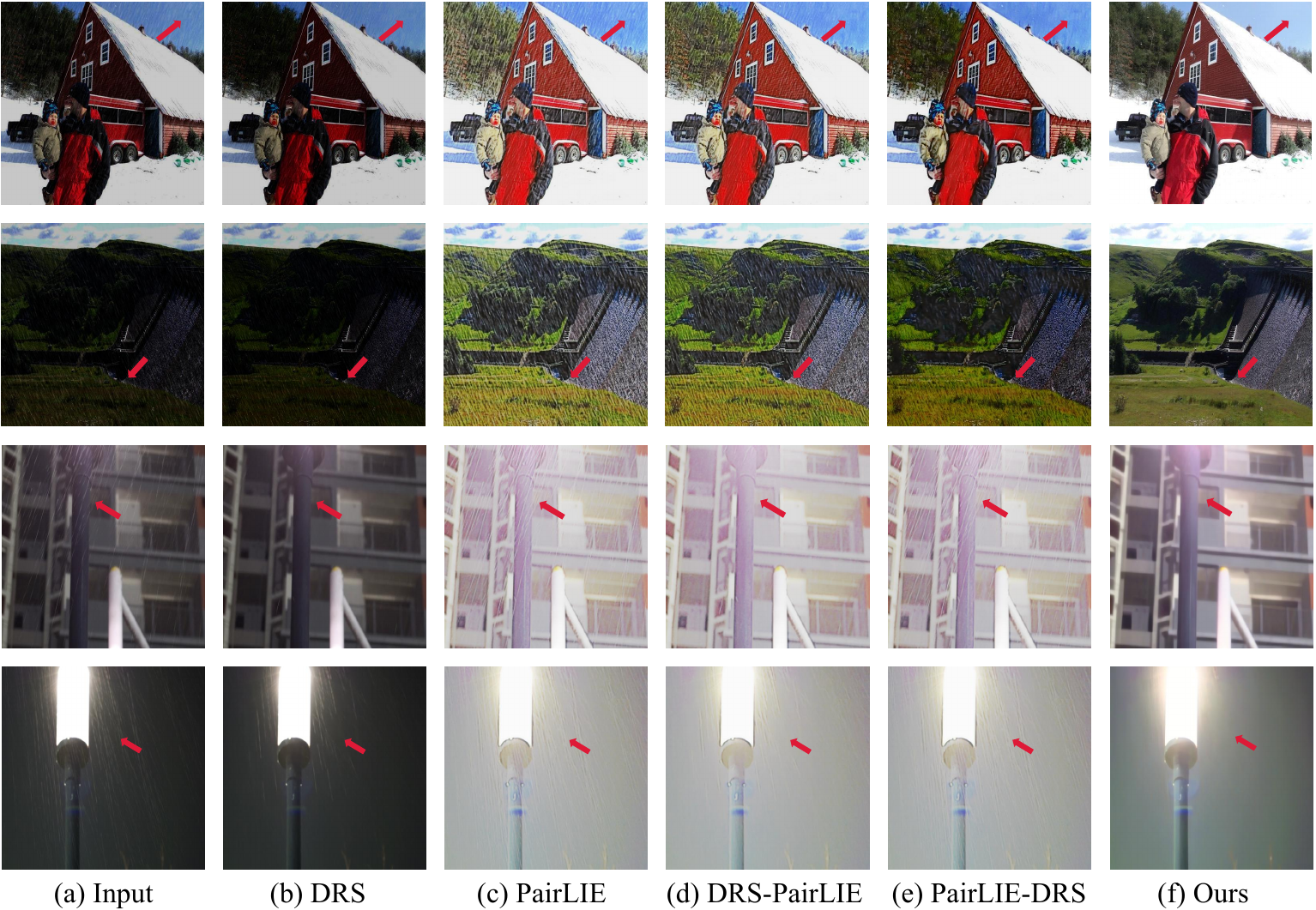}
  \caption{\textbf{Motivation.} Results from existing low-light enhancement and deraining methods on synthetic (1st and 2nd rows) and real-world (3rd and 4th rows) low-light-rainy images. (a) Input; (b) DRS \cite{drsformer}; (c) PairLIE \cite{pair}; (d) DRS \cite{drsformer} + PairLIE \cite{pair}; (e) PairLIE \cite{pair}  + DRS \cite{drsformer}; (f) L$^{2}$RIRNet (Ours). While single-task models can only address either low-light enhancement or deraining, the cascaded methods may result in issues such as overexposure, underexposure, residual rain patterns, and blurring. Our L$^{2}$RIRNet can achieve better results. The results highlight the challenges of existing approaches in effectively addressing low light and rain conditions.}
  \label{dongji}
\end{figure}

\section{Introduction}
\label{1}

Rainy nights are common weather conditions that affect many real-world activities, such as autonomous driving, surveillance systems, and photography. While existing single-image rain removal~\cite{rcd, drsformer, rlp, tcsvt_derain1, tcsvt_derain2, tcsvt_derain3, tcsvt_derain5}, low-light enhancement~\cite{zero, zero++, pair, enlightengan, tcsvt_ll1, tcsvt_ll2, tcsvt_ll5, tcsvt_ll6} and image restoration~\cite{naf, restormer, mir, scpgabnet, lrlrnet, dinoir} methods have made significant progress in addressing specific degradations, their effectiveness in solving the joint-degradation low-light-rainy task remains limited due to that rain patterns are hidden in the darkness and hard to catch.

Considering the rare work required for this joint restoration task, we conduct preliminary experiments using a cascade pipeline of two single-task models but still obtained unsatisfactory results. As illustrated in Fig.~\ref{dongji}(b), the deraining method DRS~\cite{drsformer}, trained on the Rain12600~\cite{fxy2} dataset, posing challenges in dark areas, leaving background information in these regions unavailable. In Fig. \ref{dongji}(c), the low-light enhancement approach PairLIE \cite{pair} makes rain patterns more visible but falls short of removing them. As illustrated in Figs. \ref{dongji}(d) and \ref{dongji}(e), combing DRS \cite{drsformer} and PairLIE \cite{pair} still presents challenges such as over- or under-exposure in specific areas. 
Thus, simultaneous low-light enhancement and rain removal is a non-trivial task that needs to be elaborately designed.

To address the abovementioned challenges, we analyze the physical properties of real-world low-light-rainy images and observe distinct degrading representations at different locations. The dark regions, characterized by low illumination, visually obscure rain information, while bright regions exhibit visible rain patterns. That is, different regions suffer from different types of degradation. If we handle them uniformly, it will confuse the model. Therefore, we propose an effective Low-Light-Rainy Image Restoration Network named L$^{2}$RIRNet that separately learns representations of rain streaks and low-light conditions. Moreover, we construct an LLR dataset comprising two components: synthesized image pairs and real-world images for training the model. 
The framework is designed to handle low-light-rainy scenarios by learning a hierarchical degradation representation. 
It mainly contains two key components: Dual Degradation Representation Network (DDR-Net) and Restoration-Net.

The novel two-branch representation learning DDR-Net is divided into a rainy branch, which learns rain pattern representations in bright regions, and a light branch, which learns low-light representations in dark regions. The DDRNet facilitates representation learning constrained by presented dual degradation loss (DDLoss).

Rain streaks are hidden in the darkness. Although they are invisible to the naked eye, we can detect their objective existence and significant impact on image restoration tasks after enhancing brightness with low-light enhancement algorithms. Therefore, removing rain streaks is one of the key points to the success of the low-light-rainy image restoration task. Given the high-frequency characteristics of rain patterns and their impact on the low-frequency content contour structure, we employ different channels to separate low-frequency and high-frequency information. Channel mutual reduction helps attenuate the impact of high-frequency rain patterns. Therefore, we also design a Fourier Detailed Guidance module to utilize the near-rainless detailed image's frequency and spatial domain features to guide the restoration process.

For the training, we have two stages with decoupled targets: one stage to construct the dual degradation representation of low-light-rainy images and the other stage to eliminate degradation artifacts, generating a high-quality image under the supervision of the degradation representation. Our contributions can be summarized as follows:

\begin{itemize}
	\item[$\bullet$] We propose a novel two-stage approach L$^{2}$RIRNet to achieve low-light-rainy image restoration in synthetic and real scenes. Experimental results have proven the effectiveness of our L$^{2}$RIRNet.
	\item[$\bullet$] We develop a DDR-Net containing rainy branch and light branch encoders to learn different degradation representations separately. This design provides better guidance than a single one for the restoration process.
        \item[$\bullet$] We present an FDG module that combines spatial and frequency domain information of our designed near-rainless detailed images to provide effective prior for the restoration part.
\end{itemize}

\begin{figure*}[ht]
\includegraphics[width=1\linewidth]{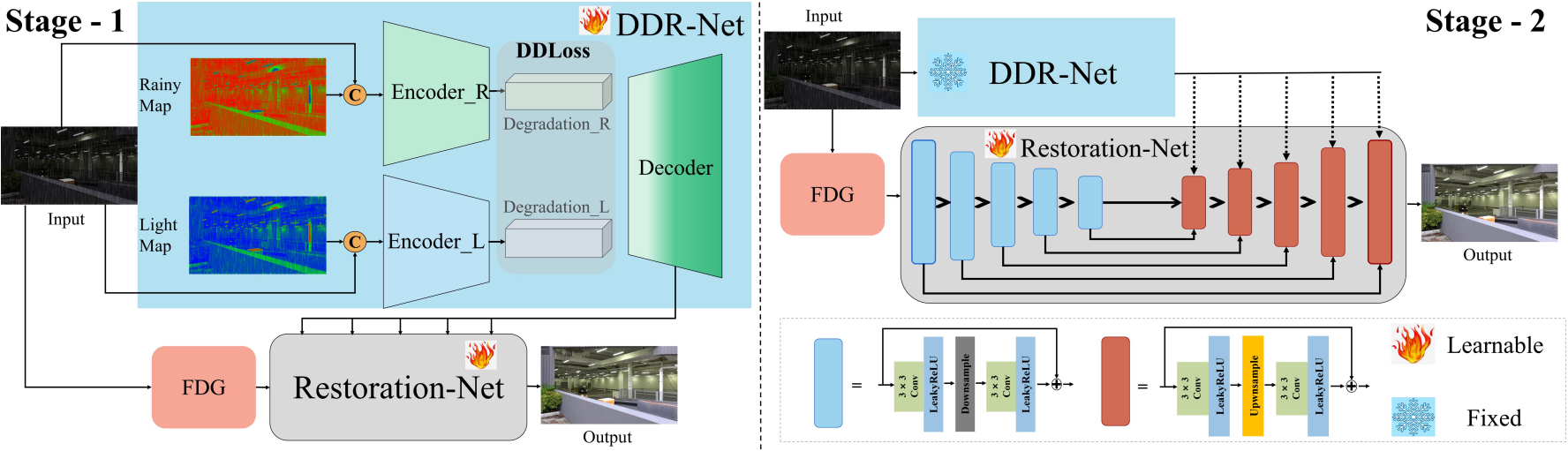}
\caption{\label{kuangjia}
Overview of our L$^{2}$RIRNet consists of two components: DDR-Net and Restoration-Net. DDR-Net extracts the image's degradation representation of bright and dark regions by rainy and light maps, respectively. The dual degradation loss (DDLoss) is presented to constrain the training process. The decoder generates the multi-channel latent features to guide the restoration process effectively. The Fourier Detailed Guidance (FDG) module uses prior near-rainless information for the Restoration-Net. The training process consists of two phases: 1. Train both DDR-Net and Restoration-Net to get effective representations. 2. Fix the DDR-Net and only train the Restoration-Net to enhance restoration performance further.}
\end{figure*}

\section{Related Work}

\subsection{Single Image Rain Removal.}

Image deraining poses a significant challenge due to its inherently ill-posed nature. Traditional methods reconstruct the background from rainy images by extracting the rain-free high-frequency component (HFP). Various filtering strategies, such as guided filters \cite{filter1}, multiple guided filtering \cite{filter3}, and nonlocal means filtering \cite{filter5}, have been employed for this purpose. 
For the deep learning-based methods. In \cite{fxy1}, the proposed CNN-based framework can extract discriminative rain features from the HFP of a single rainy image.
A physics-based model is introduced in \cite{tip1}, which resembles real rain scenes, as an initial constraint for deraining. A rain-density classifier guides subsequent refinement. 
On the other hand, PFDN \cite{tip2} is a novel image derivation framework employing rain-related and rain-independent features to eliminate rain streaks and recover contextual details from multiple perspectives. 
However, the paired datasets are difficult to obtain in the real world, so a Gan-based unsupervised deraining framework is proposed in \cite{tcsvt_derain3}.
In \cite{tcsvt_derain5}, the authors create a large rainy database to enhance deraining performance.
The deraining model framework is also one of the directions of efforts in method design. For instance, a multi-scale architecture and attention strategy are presented in \cite{tcsvt_derain2}. 
Other approaches, such as \cite{sgi}, construct a new research direction that explores the use of high-level semantic information to enhance rain removal. 
Additionally, researchers address more complex task settings, such as synchronous rain streaks and raindrop removal in \cite{78}. 
In recent advancements, MIRNet \cite{mir} uses a novel feature extraction network effective in image restoration and enhancement. 
On the other hand, RCDNet \cite{rcd} introduces an unfolding method, employing multi-stage training with M-net and B-net, achieving superior performance in image deraining. 
Recently, Restormer is introduced in \cite{restormer}, an approach utilizing transformers for deraining, capitalizing on transformer benefits while minimizing computational complexity. 
NAFNet \cite{naf} employs a sequence of uncomplicated yet impactful improvements, optimizing the network baseline and fully unlocking its performance advantages. 
In \cite{drsformer}, a multi-experts-based DRSFormer provides more accurate detail and texture recovery. 
Most recently, RLP \cite{rlp} points out a nighttime deraining method emphasizing the significance of incorporating location information regarding rain streaks, especially in night scenes. 
However, all these methods are tailored for restoring rain images captured in daytime conditions and may struggle to remove rain patterns effectively in low-light conditions.

\subsection{Low-light Image Enhancement.}
Traditional methods employ handcrafted optimization techniques and norm minimization methods to enhance low-light images. These approaches rely on various assumptions, such as dark/bright channel priors \cite{e11, e10, e12}, and mathematical models, including Retinex theory \cite{e6} and Multi-scale Retinex theory \cite{e4, e5}, to derive sub-optimal solutions. Learning-based methods fall into four categories: supervised \cite{e10}, self-supervised \cite{sci}, unsupervised \cite{enlightengan}, and zero-shot \cite{zero}. Kind \cite{kind} considers the impact of noise in low-light images and utilizes a noise removal module to enhance performance. Zero-shot approaches suggest various approximation strategies to reduce label dependency. Zero-DCE \cite{zero}, and Zero-DCE++ \cite{zero++} estimate multiple tone curves from input images to obtain enhanced results. A physical model is proposed in \cite{tcsvt_ll5}, which uses weight coefficients based on the light-scattering rate.
Various low-light conditions need to be considered \cite{tcsvt_ll2}; the authors propose a method to handle multiple situations to improve the network's generalization.
In \cite{tcsvt_ll1}, the authors present a unified Retinex-based zero-reference deep framework with an effective generative strategy.
On the other hand, a novel self-supervised framework SCI \cite{sci} is introduced with a cascaded illumination learning approach to minimize redundancy and enhance overall efficiency. 
PyDiff \cite{ijcai} is the first to consider the diffusion model in low-light enhancement, achieving superior performance and speed compared to existing methods. Recently, a paired-based low-light enhancement technology PairLIE \cite{pair} with designed priors and a simplified network. In \cite{uhd}, UHDNet is introduced, which draws inspiration from distinctive features identified in the Fourier domain. 
However, for low-light-rainy images, these methods only enhance brightness and make rain patterns more visible in dark areas, and they can not effectively restore them to a high-quality version.

\begin{figure}[h]
\centering
\includegraphics[width=1\linewidth]{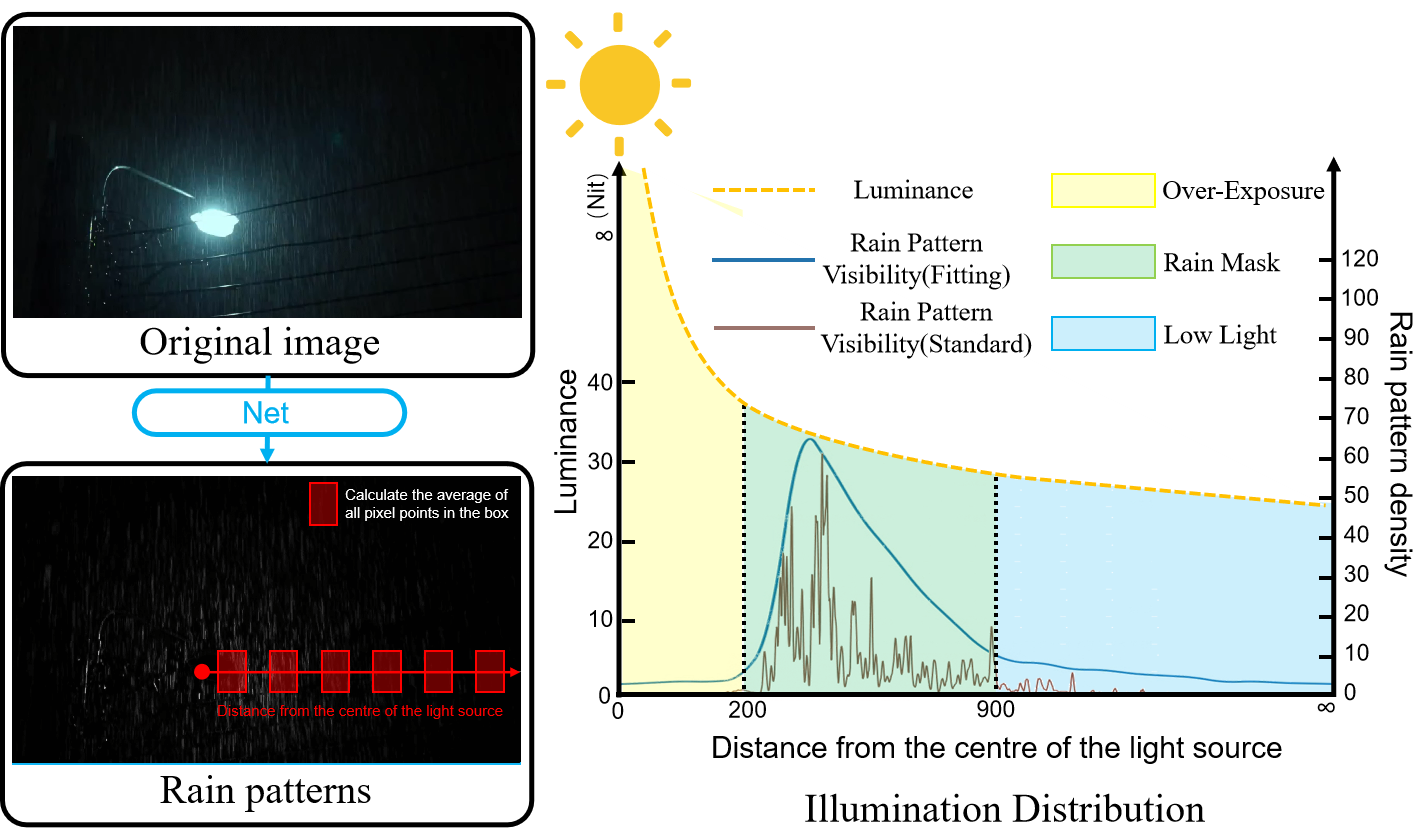}
\caption{\label{map}
The illustration of the relationship between rain pattern density (m) and the Euclidean distance (r) from the pixel to the light source.}
\end{figure}

\section{Proposed Method}

\subsection{Overview}
Our L$^{2}$RIRNet consists of two key components: a Dual Degradation Representation Network (DDR-Net) and a Restoration Network. As depicted in Fig. \ref{kuangjia}, we utilize rainy and light attention maps to learn distinct degradation representations separately. The representation learning process is optimized through dual degradation contrastive loss (DDLoss). Our objective is to capture information that significantly influences the image quality of low-light-rainy images. The DDR-Net encodes the degradation operator into a low-dimensional feature to guide the restoration process, employing a U-Net-like network structure. Overall, the network is trained in two stages. We train the DDR-Net and Restoration-Net in the first stage for more accurate representations. In the second stage, we lock the DDR-Net and only train the Restoration-Net to improve the restoration performance further.

In the case of the DDR-Net, we initiate the process by acquiring a fusion degradation representation through $Att\_R$ and $Att\_L$. Subsequently, we derive the latent degradation representation $\Theta\_Latent$ using the Decoder:
\begin{equation}
\left\{
\label{gonshi}
\begin{array}{l}
Att\_R = Concat[Input, Map\_R], \\
Att\_L = Concat[Input, Map\_L], \\
\Theta\_R = \Phi_R(Att\_R), \\
\Theta\_L = \Phi_L(Att\_L), \\
\Theta\_F = Concat[\Theta\_R, \Theta\_L], \\
\Theta\_Latent = Decoder(\Theta\_F),
\end{array}
\right.
\end{equation}
where $\Phi_R(\bullet)$ and $\Phi_L(\bullet)$ denote the encoder in rainy and light branch, respectively. $\Theta$ is the degradation representation. 
Subsequently, the Restoration-Net utilizes the $\Theta\_Latent$ acquired from the DDR-Net to guide the restoration process, resulting in the final output:
\begin{equation}
\begin{split}
\label{gonshi_2}
Output = Restoration(Input, \Theta\_Latent). 
\end{split} 
\end{equation}

Finally, the restored image is obtained. The key components are elaborated below.

\subsection{Dual Degradation Representation Network (DDR-Net)}
\label{3.2}
\subsubsection{Adaptive Map for Low-light-rainy Images}
\label{3.2.1}

As depicted in Fig. \ref{map}, we analyze the relationship between light intensity and rain pattern density using a real-world low-light-rainy image containing a light source. We set $E$ as light intensity and $r$ as the Euclidean distance from surrounding pixels ($x_1$, $y_1$) to the center ($x_0$, $y_0$) of a light source, expressed as $r = \sqrt{(x_1-x_0)^2 + (y_1-y_0)^2}$. The physical relationship curve between $E$ and $r$ as $E = \frac{E_0}{r^2}$, where $E_0$ represents the light intensity at the center of the light source. To calculate the rain pattern density $m$, we utilize a pre-trained RCDNet \cite{rcd} to obtain the rain pattern mask. Then, we generate numerous 20$\times$20 patch blocks intersecting the straight line. Finally, the relationship between $m$ and $r$ is shown in the right part of Fig. \ref{map}.

Especially in the pixel region between 0 and 200, capturing raindrop information is challenging due to overexposure near the light source; between 200 and 900, raindrop information becomes more visible, and the degraded information is predominantly composed of raindrop details; and in the region between 900 and infinity, capturing raindrop information is also difficult due to image degradation caused by low-light conditions.

In Fig. \ref{kuangjia}, we employ the attention map proposed by \cite{enlightengan} to accentuate the dark areas of the image, creating our Light Map. This allows the network to capture information related to low-light degradation effectively. Simultaneously, by utilizing the attention map's complement, the image's bright regions correspond to the rain pattern information in those areas, forming our Rainy Map. By training the network to independently process the bright and dark areas using the $Map\_R$ and $Map\_L$, we can derive the degradation latent and effectively guide the Restoration-Net.

\begin{figure*}[t]
\centering
\includegraphics[width=1\linewidth]{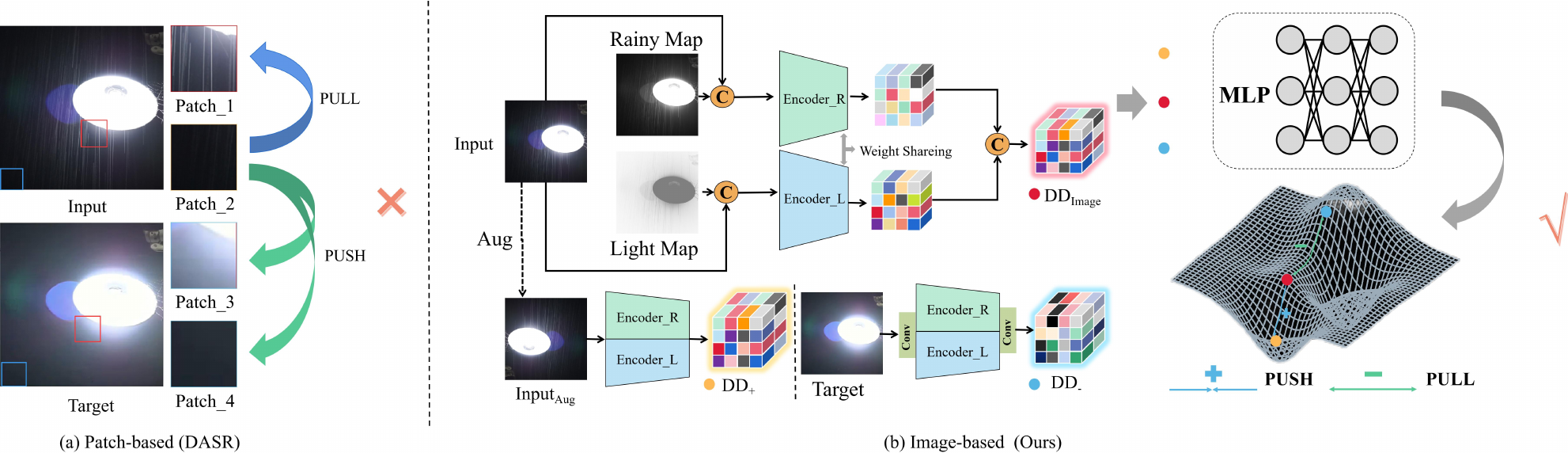}
\caption{\label{pdie_1} (a) DASR's patch-based positive and negative label approach. The selected patches may come from low-light and rain pattern areas, which is challenging to apply as a positive label. (b) Our image-based method learns degradation representation by the dual degradation loss. We present the rainy and light encoder part of the DDR-Net, which takes in three images: a low-light-rainy image (Input), an augmented version of that image ($Input_{aug}$), and a paired clean image (Target). These images are encoded using a parameter-sharing network, with rainy and light maps focusing on the degradation representation in bright and dark regions. A multilayer perceptron (MLP) then processes the resulting feature information to obtain the degradation operator for comparison learning constraint. The $DD_{Image}$ is the dual degradation representation from Input, the $DD_{+}$ is the dual degradation representation from $Input_{aug}$, and the $DD_{-}$ is the dual degradation representation from Target. Our approach aims to make $DD_{Image}$ and $DD_{+}$ similar while keeping $DD$ and $DD_{-}$ distinct.
}
\end{figure*}

\begin{figure}[ht]
\centering
\includegraphics[width=1\linewidth]{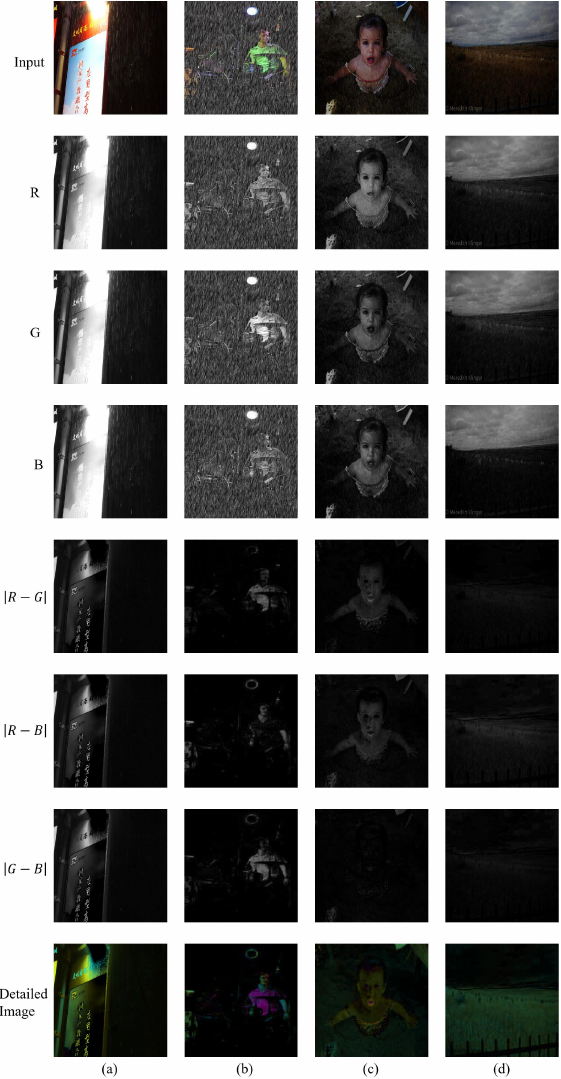}
\caption{\label{detail} Visual presentation of detailed images. Compared to the input, the detailed image removes most raindrops while maintaining the content information.
}
\end{figure}

\begin{figure*}[ht]
\centering
\includegraphics[width=1\linewidth]{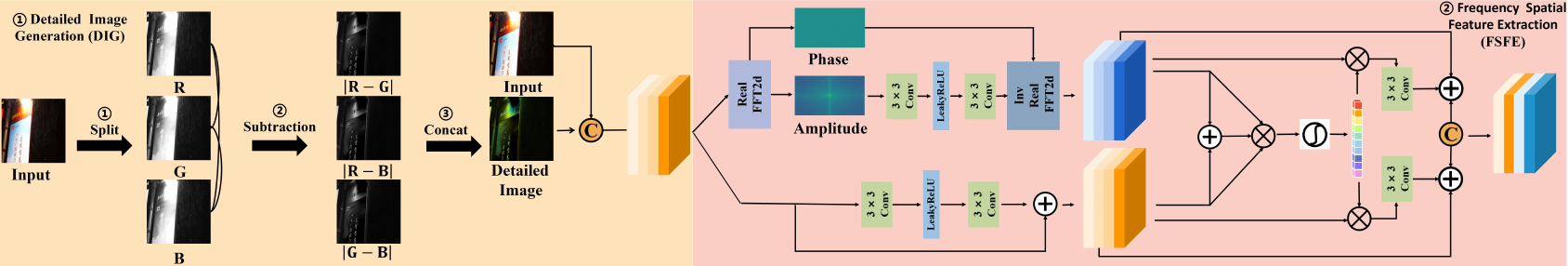}
\caption{\label{dgm}
The architecture of the FDG module is as follows, comprising two main steps: 1. Detailed Image Generation (DIG): initially splits an input low-light-rainy image into RGB channels. Subsequently, it subtracts the absolute differences between these channels, creating $\mathrm{\left| R-G \right|}$, $\mathrm{\left| R-B \right|}$ and $\mathrm{\left| G-B \right|}$. Finally, concatenates them to form a detailed image. 2. Frequency Spatial Feature Extraction (FSFE). FSFE contains two branches: the frequency branch and the spatial branch. The frequency branch focuses on capturing frequency-related information, while the spatial branch is designed to extract spatial features. The resulting fusion of these features is employed to guide the restoration process effectively.}
\end{figure*}

\subsubsection{Dual Degradation Contrastive Loss}

Utilizing contrastive learning loss (CL) can not only significantly enhance the performance \cite{duibiquwu} but also extract and learn degradation representation to guide the downstream process \cite{dasr, face, degra}. In such cases, the mixed degradation representation can be expressed using a multiplication or addition formula.

\textbf{Limitation of CL.} One of the most classic methods is the contrastive learning scheme of DASR \cite{dasr}, which selects two patches from the same image as the anchor and the positive label and lets patches from the paired clean image be the negative labels. However, our previous analysis in section \ref{1} and \ref{3.2.1} reveals significant differences in rainy and dark levels across various positions in a single low-light-rainy image, as depicted in Fig. \ref{pdie_1}(a). This makes the patch-based approach unsuitable for low-light-rainy image restoration tasks. Thus, in the next section, we introduce different degradation representations of the whole image separately.

\textbf{Solution.} To solve the limitations above, we present a dual degradation loss (DDLoss). As shown in Fig. \ref{pdie_1}(b), firstly, we abstain from learning the mixed degradation representation and instead concentrate on autonomously acquiring the representations for rainy and light degradation. Following this, we proceed to integrate these representations. Secondly, rather than selecting the anchor and the positive label from the same image, we propose an alternative approach that incorporates a paired clean Target ($x_{tar}$) as the negative label and the augmented version of Input($x$), named $\mathrm{Input_{Aug}}$ ($x_{aug}$) (rotations of 90, 180, and 270 degrees and horizontal flipping), as the positive label. The fused degradation from the DDR-Net encoder ($DDR_{E}$) is input to a two-layer multi-layer perceptron (MLP) projection head to generate three vectors. The learned degradation allows anchor patches to better approach ``positive'' pairs in a specific metric space and move away from ``negative'' pairs. The dual degradation loss (DDLoss) function is defined as:

\begin{equation}
\begin{split}
L_{D}=\dfrac{\Vert MLP(DDR_{E}(x))-MLP(DDR_{E}(x_{aug}))\Vert_{1}}{\Vert MLP(DDR_{E}(x))-MLP(DDR_{E}(x_{tar}))\Vert_{1}}.
\end{split} 
\end{equation}

\subsubsection{DDR-Net Framework}

Based on the discussion above, we propose an effective DDR-Net with a U-Net-like structure for extracting degradation information. It includes components for attention map extraction, an encoder, and a decoder. By utilizing rainy and light maps, we divide the image into bright and dark regions, directing the network's attention more intensively to each area during training. Specifically, we extract distinct degradation information from the two regions and fuse them.

\subsection{Restoration-Net}

\subsubsection{Fourier Detail Guidance Module (FDG)}

Low illumination and noise affect not only image content but also the distribution of rain patterns, thereby impacting image visibility. Removing the rain pattern information while recovering the lighting is challenging for the restoration part. We observe that the contents of the rain streaks in three RGB channel maps of the input image are similar. Thus, near-rainless detailed images can be obtained by calculating the difference between these channels, as shown in Fig. \ref{detail}. Thus, we leverage this characteristic to reduce rain patterns' interference and deeply explore the image content information. Specifically, we implement a Fourier Detail Guidance (FDG) module. This module utilizes a detailed image of the background without rain patterns to provide well-defined prior information for the restoration process. As shown in Fig. \ref{dgm}, the FDG contains two components: Detailed Image Generation (DIG) and Frequency Spatial Feature Extraction (FSFE). 

The input low-light-rainy image is initially split into the RGB (red, green, and blue) channels in the DIG step. Subsequently, the absolute differences between each pair of channels are calculated, resulting in three single-channel maps: $\mathrm{\left| R-G \right|}$, $\mathrm{\left| R-B \right|}$ and $\mathrm{\left| G-B \right|}$. Finally, these maps are concatenated to create a detailed image that captures the unique features of each single-layer map. Fig. \ref{detail} illustrates that the detailed image for both the synthetic and the real-world image obtained by DIG preserves the image's detailed structure effectively and removes the rain pattern information. This approach serves as a preliminary rain removal strategy, and the resulting images can be effectively utilized for the restoration tasks. 

To further leverage the detailed images in guiding the restoration process, we design FSFE. The detailed image is passed to parallel frequency/spatial domain branches, each featuring two distinct paths for processing input feature channels: a spatial path applying conventional convolution to a subset of channels and a frequency path operating in the Fourier domain. The Fourier transform is a widely used tool for analyzing the frequency content of signals. This allows for the efficient analysis and processing of images and other complex signals using Fourier-based techniques. Utilizing element multiplication, we enhance the detailed image and then capture the resulting features to obtain prior information to guide the preliminary restoration of the image. Our experimental results demonstrate that the proposed method effectively removes rain patterns from the image while preserving background content information. As a result, the FDG module can concentrate on learning the background information, which enhances its ability to remove rain pattern information.
\label{3.3.1}

\subsubsection{Restoration-Net Framework}
Our Restoration-Net architecture is specifically designed to restore low-light-rainy images to their corresponding clean versions. This is achieved through the FDG module for pre-processing, an encoder, and a decoder. 

As detailed in the previous section, the FDG module processes both the detail image and the low-light-rainy image, concatenating them to form the encoder's input. This pre-processing step can reduce noise and rain patterns, preserve image details, and enhance contrast. The encoder is responsible for extracting feature information from the input image by increasing its dimension. The encoded feature information is transmitted to the decoder via a mid-level jump connection.

To improve the performance of the restoration process, the Restoration-Net decoder leverages the degradation feature from multiple layers by DDR-Net. This strategic utilization enables the Restoration-Net architecture to concentrate more on the degraded components of the image, thereby improving the efficiency and performance of the restoration process. 

We use the L1 loss and perceptual loss to train our Restoration-Net. Specifically, the loss function is defined as:
\begin{equation}
\begin{split}
\label{r}
L_{R}= \Vert R(x)-x_{clean}\Vert_{1} + \lambda_{p}\Vert \Phi(R(x)) - \Phi(x_{clean})\Vert_{1},
\end{split} 
\end{equation} 
where $R(\bullet)$ denotes the Restoration-Net, $\lambda_{p}$ is the weight parameters, $\Phi(\bullet)$ denotes the pre-trained VGG19 network and we adopt multi-scale feature maps from layer $\left\{ conv1, ..., conv4\right\}$ following the widely-used setting. 

The overall loss function is:
\begin{equation}
\begin{split}
\label{L}
L= \lambda_{D}L_{D} + \lambda_{R}L_{R},
\end{split} 
\end{equation} 
where $\lambda_{D}$ and $\lambda_{R}$ denote weight parameters.

\begin{table*}
\begin{center} 
\fontsize{10}{10}\selectfont
\caption{Quantitative evaluation on the LLR Dataset. The baseline contains the cascaded methods (Enhancement-Deraining and Deraining-Enhancement) and retrained methods on our LLR Dataset. `*' indicates the network is retrained on our LLR Dataset. The parameters and running time are expressed separately in millions (M) and milliseconds (Ms). All running time is evaluated on a 512 $\times$ 512 image using a Geforce RTX 3090.} 
\setlength{\tabcolsep}{4.3mm}{
\begin{tabular}{cccccc}
\toprule
& Methods & PSNR(dB) & SSIM & Param(M) & Running time (Ms)\\
\midrule
\myrowcolour%
\multirow{10}{*}{Cacaded Methods} & Zero-DCE $\Rightarrow$ RCDNet & 17.58& 0.6634 & 3.5& 235.33\\

\cmidrule{2-6}
& SCI $\Rightarrow$ MIRNet & 20.71& 0.7186 & 32.3 & 198.50\\
\cmidrule{2-6}
\myrowcolour%

& UHDNet $\Rightarrow$ Restormer & 19.91 & 0.6928 & 59.8 & 206.64\\
\cmidrule{2-6}

& PairLIE $\Rightarrow$ DRSFormer & 18.29& 0.6495 & 34.0& 555.38\\
\cmidrule{2-6}
\myrowcolour%

& MIRNet $\Rightarrow$ SCI & 19.32& 0.6931 & 32.3& 198.50\\
\cmidrule{2-6}

& RCDNet $\Rightarrow$ Zero-DCE & 12.21& 0.6175 & 3.5& 235.33\\
\cmidrule{2-6}
\myrowcolour%

& Restormer $\Rightarrow$ UHDNet & 19.57&0.6878 & 59.8 & 206.64\\
\cmidrule{2-6}

& DRSFormer $\Rightarrow$ PairLIE & 19.19& 0.6868 & 34.0&555.38\\

\midrule

\myrowcolour%
\multirow{16}{*}{Retrained Methods} & Zero-DCE* & 21.75& 0.7189 & 0.5& 57.81\\

\cmidrule{2-6}
& RCDNet* & 25.04& 0.8031 & 3.0& 177.52\\
\cmidrule{2-6}
\myrowcolour%

& MIRNet* & 29.31& 0.8872 & 31.8& 193.53\\
\cmidrule{2-6}

& Restormer* & 31.19& 0.9174 & 26.1 & 190.53\\
\cmidrule{2-6}
\myrowcolour%

& NAFNet* & 29.17& 0.8987 & 17.1& 32.67\\
\cmidrule{2-6}

& PyDiff* & 29.06& 0.8753 & 54.5&373.56\\
\cmidrule{2-6}
\myrowcolour%

& UHDNet*& 24.91& 0.7914 & 33.6& 16.11\\
\cmidrule{2-6}

& PairLIE*& 19.33& 0.6948 & 0.3&2.42\\
\cmidrule{2-6}
\myrowcolour%

& RLP* & 29.45& 0.8878 & 5.3& 101.72\\
\cmidrule{2-6}

& DRSformer* & 31.58& 0.9182 & 33.7& 552.96\\
\cmidrule{2-6}
\myrowcolour%

& L$^{2}$RIRNet-CNN (ours) & 30.53& 0.9030 & 73.9& 58.38\\
\cmidrule{2-6}

& \color{red}{L$^{2}$RIRNet-Trans (ours)} & \color{red}{32.53} & \color{red}{0.9231} & 37.5& 229.01\\

\midrule
\end{tabular}}%
\label{table1}
\end{center} 
\end{table*}%

\begin{figure}[t]
\centering
\includegraphics[width=1\linewidth]{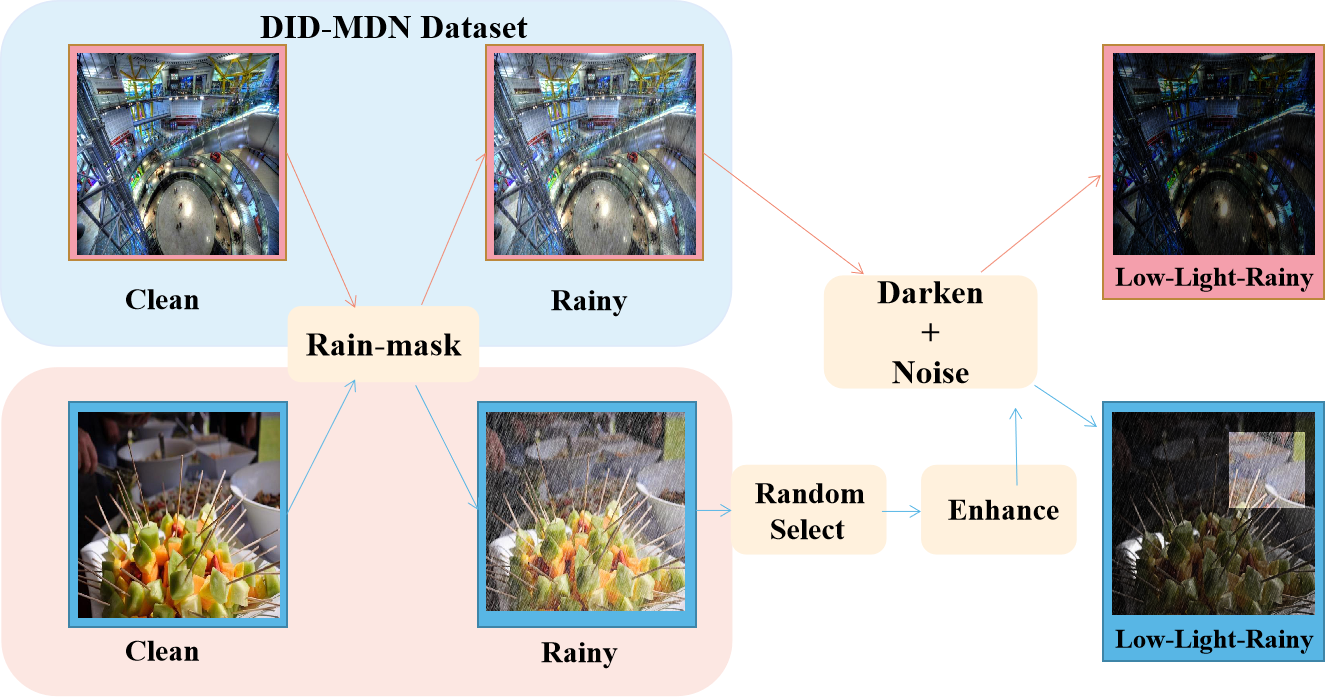}
\caption{\label{data}
An overview of synthetic images in the LLR dataset synthesis pipeline. The clean and rainy images belong to the DID-MDN dataset \cite{rain12000}. To obtain (1), we darken and add noise to 4000 images. For (2), we randomly select patches to enhance the light intensity and darken other regions, simulating real-world low-light-rainy images.}
\end{figure}

\begin{figure}[t]
\centering
\includegraphics[width=1\linewidth]{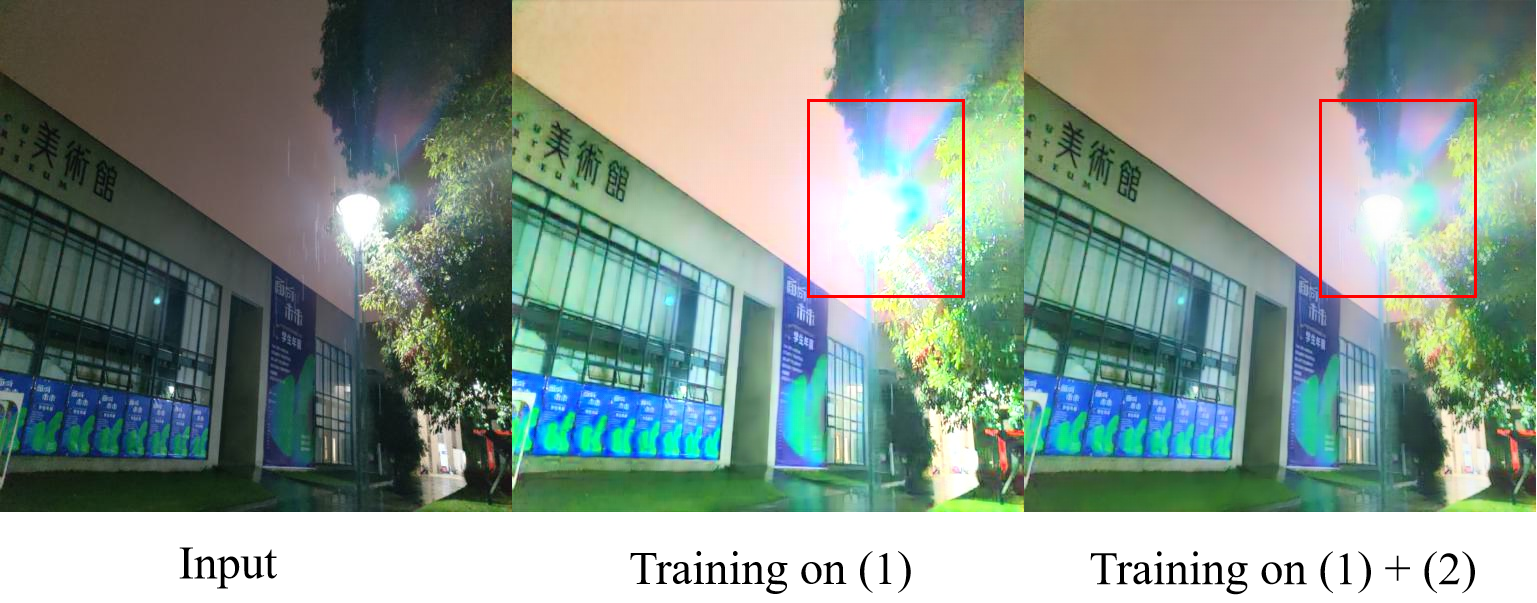}
\caption{\label{trick}
A visual comparison on a real-world image of our L$^{2}$RIRNet training only on (1) and (1) + (2).}
\end{figure}

\begin{figure}[t]
\centering
\includegraphics[width=1\linewidth]{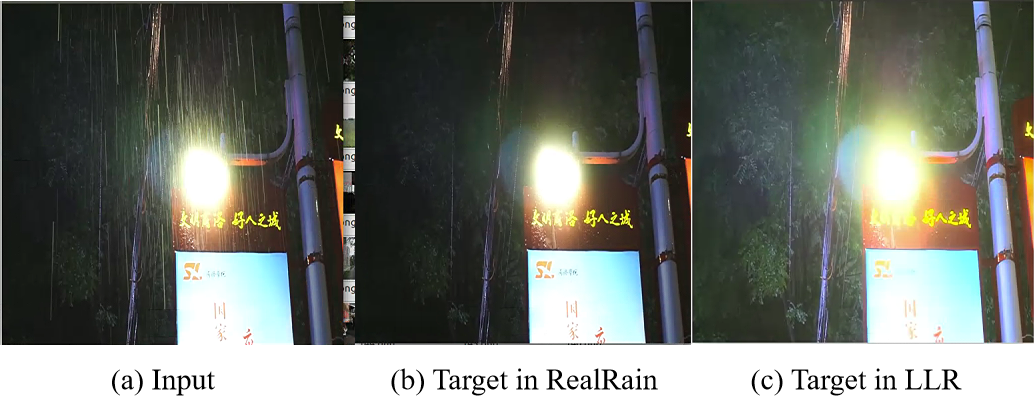}
\caption{\label{realrain}
(a) and (b) are a real rainy image and its clean target from \cite{data1}. (c) is obtained by enhancing the light of (b) as the target of the low-light-rainy restoration task.}
\end{figure}

\begin{figure*}[ht]
\centering
\includegraphics[width=1\linewidth]{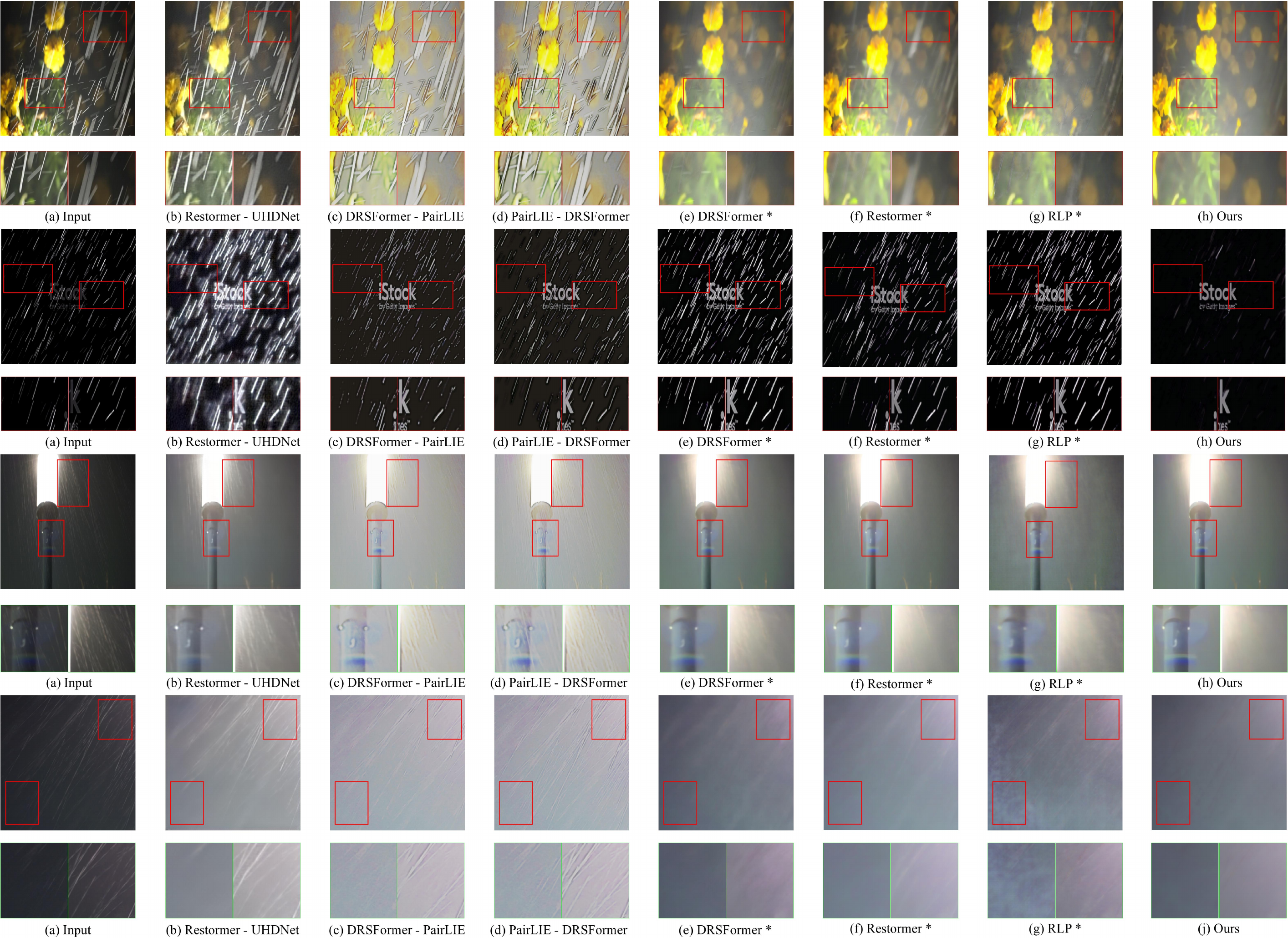}
\caption{\label{zhenshi_1} Visual comparison on a real-world image from the LLR dataset. ‘*' indicates the network is retrained on our LLR Dataset.
}
\end{figure*}

\begin{figure}[ht]
\centering
\includegraphics[width=1\linewidth]{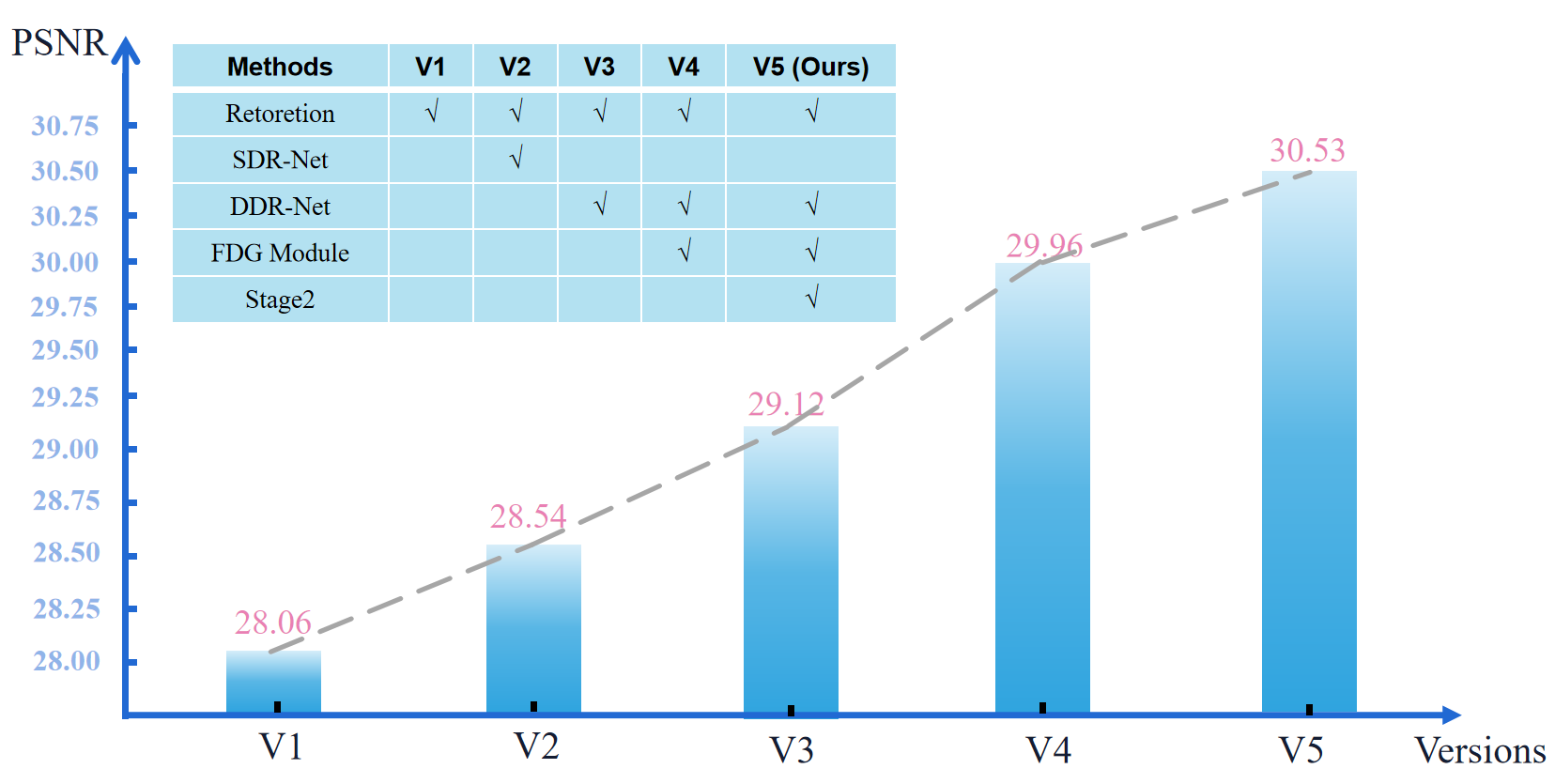}
\caption{\label{xiaorong} $\textbf{V1}$: Only the Restoration-Net; $\textbf{V2}$: Restoration-Net + SDR-Net (single degradation representation network) ; $\textbf{V3}$: Restoration-Net + DDR-Net (dual degradation representation network); $\textbf{V4}$: The first stage of L$^{2}$RIRNet: V3 + Fourier Detailed Guidance (FDG) module. $\textbf{V5 (ours)}$: V4 + second stage of L$^{2}$RIRNet.
}
\end{figure}

\begin{table*}
\begin{center} 
\fontsize{10}{10}\selectfont
\caption{Quantitative evaluation on the real-world LLR Dataset.} 
\setlength{\tabcolsep}{6.8mm}{
\begin{tabular}{ccccccc}
\toprule
\myrowcolour%
& Input & RCDNet* & MIRNet* & Restormer* & NAFNet* \\
\midrule
NIQE $\downarrow$ & 16.93 & 15.97 & 14.31& 15.55& 14.65\\
\midrule
\myrowcolour%
CEIQ $\uparrow$ & 3.31 & 3.36& 3.43& 3.47& 3.46 \\
\midrule
& DRSformer*& RLP*& UHDNet*& PairLIE*& \color{red}{L$^{2}$RIRNet (Ours)} \\
\midrule
\myrowcolour%
NIQE $\downarrow$ & 14.82 & 14.61 & 16.36& 17.34& \color{red}{14.17} \\
\midrule
CEIQ $\uparrow$ & 3.45 & 3.44 & 3.49& 3.36& \color{red}{3.50} \\
\midrule

\end{tabular}}%
\label{table5}
\end{center} 
\end{table*}%

\begin{figure*}[ht]
\centering
\includegraphics[width=1\linewidth]{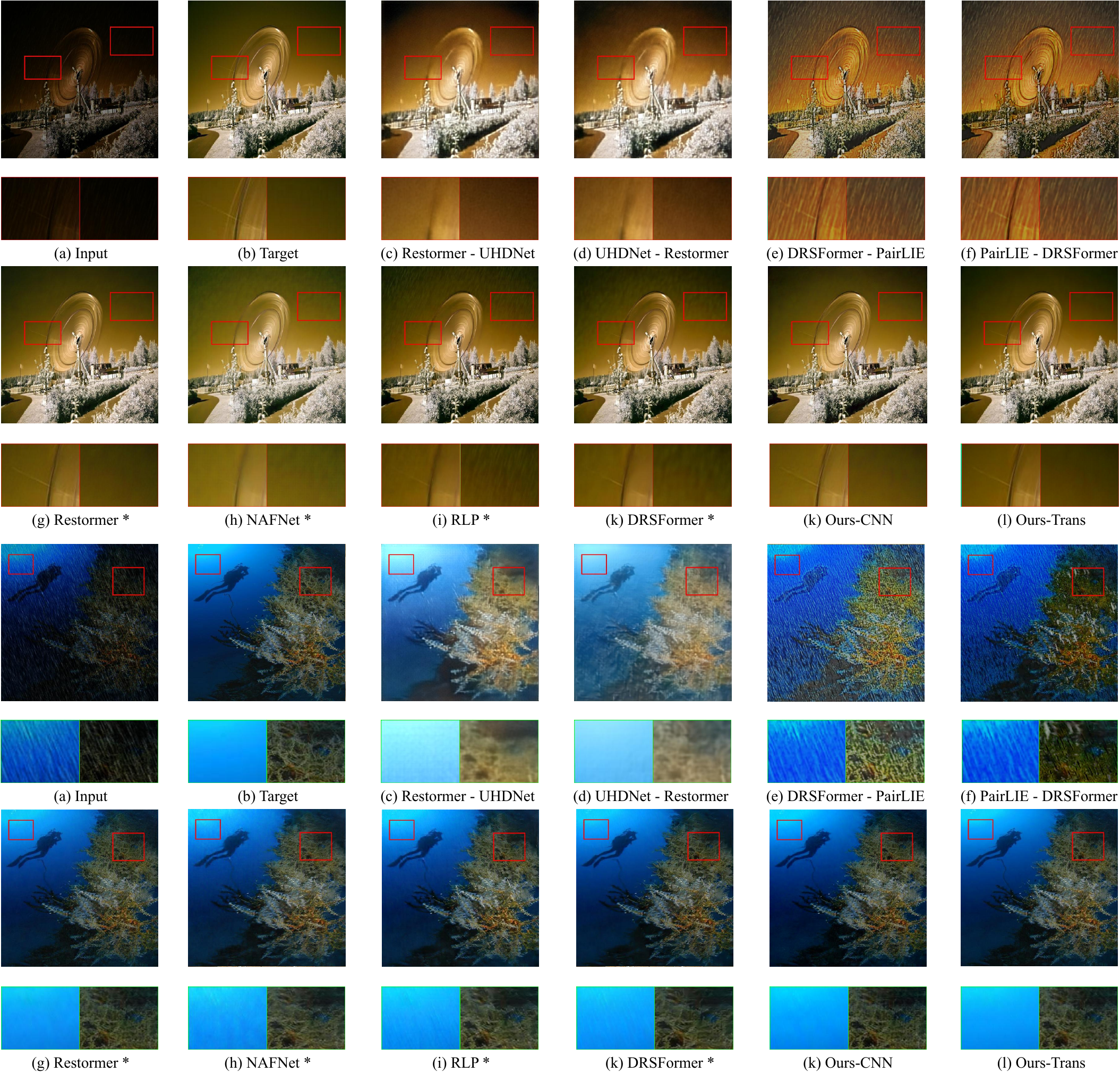}
\caption{\label{hecheng} Visual comparison on synthetic images from the LLR dataset. ‘*' indicates the network is retrained on our LLR Dataset.
}
\end{figure*}

\section{Experiments}


\subsection{Proposed Dataset} 

For training the proposed L$^{2}$RIRNet, we contribute an LLR Dataset containing synthetic and real-world images. It contains 8368 pairs of images for training, 800 synthetic, and 42 real-world images for testing.

\noindent \textbf{Synthetic low-light-rainy images}: 

(1) Inspired by LEDNet \cite{led}, which proposes a pipeline for generating low-light blurred images, we first synthesize 8000 low-light-rainy images by decreasing the lighting intensity of synthetic rainy images from the DID dataset \cite{rain12000}. Then, we add random Gaussian noise to simulate real-world low-light conditions.

(2) Considering the lighting distribution characteristics of real-world environments, we randomly select 4000 images from the first part and enhance the pixels within random patches while reducing the lighting in other areas. The specific steps are illustrated in Fig. \ref{data}.

As shown in Fig. \ref{trick}, we present the results of a real-world low-light-rainy image. Our L$^{2}$RIRNet is trained on the dataset (1) and (1) + (2), respectively. The addition of (2) improves the resolution of the overexposed areas near the light source, demonstrating the usefulness of this operation.

We also research 368 real-world low-light-rainy images from \cite{data1} to construct fake-synthetic low-light-rainy images for training. In Fig. \ref{realrain}, (a) and (b) are a pair of rainy-clean images from \cite{data1}, and we use SCI \cite{sci} to enhance the target image to obtain (c) as our target in low-light-enhancement.

\noindent \textbf{Real-world low-light-rainy images}:
 
(1) We search the real deraining datasets \cite{data1}, \cite{data2} for nighttime scenes.

(2) We conduct a Google search to find real-world low-light-rainy images.

(3) We capture some low-light-rainy images in outdoor scenes.

\subsection{Experimental settings}

\subsubsection{Implementation details} 

Our DDR-Net and Restoration-Net are both U-Net-like architectures. We give two versions for Restoration-Net: CNN-based (MPRNet \cite{mprnet}) and Trans-based (Restormer \cite{restormer}). The scale of the DDR-Net will be matched accordingly. To optimize our proposed model, we adopt the Adam optimizer algorithm with $\beta_1$=0.9, $\beta_2$=0.999, and the initial learning rate is set to 2 $\times$ $10^{-4}$ and gradually decreases to $10^{-6}$ with the cosine annealing schedule \cite{cos}, the used framework is PyTorch. The used GPU is GeForce RTX 3090. For the CNN-based version, the batch size is 24, and the patch size is 256, at the first stage for 200 epochs and the second stage for 300 epochs. For the Trans-based version, the batch size is set to 8, and the patch size is 128, at the first stage for 90 and the second stage for 200 epochs. We set the weight weights $\lambda_{p}$, $\lambda_{D}$ and $\lambda_{R}$ in~\eqref{r} and \eqref{L}  to 0.1, 1 and 1, respectively.

\subsubsection{Evaluation Metrics} We employ the PSNR and SSIM metrics to evaluate the restored results on synthetic images in the LLR Dataset, and for comprehensive comparison, the model parameters and running time are given. For real-world images from the LLR Dataset, since there is no ground truth for calculating PSNR and SSIM, the classical non-reference NIQE \cite{niqe} and CEIQ \cite{ceiq} are used.

\subsection{Comparisons on the LLR Dataset} We evaluate the effectiveness of the proposed L$^{2}$RIRNet on the synthetic and real-world images from the LLR dataset quantitatively and qualitatively. As the joint task addressed in this paper is new, no open existing methods can directly compare with ours. To provide a comprehensive evaluation, we carefully select some representative low-light enhancement and deraining methods, resulting in two distinct baseline methods for comparison: cascaded and retrained.

\noindent
 \textbf{Cascaded Methods.}
For low-light enhancement methods, we choose the classic Zero-DCE \cite{zero}, SCI \cite{sci}, and some state-of-the-art approaches UHDNet \cite{uhd} and PairLIE \cite{pair}. For the deraining part, we choose some recent deraining methods: MIRNet \cite{mir}, RCDNet \cite{rcd}, Restormer \cite{restormer}, and DRSFormer~\cite{drsformer}.

\noindent
 \textbf{Retrained Methods.} 
In addition to the methods mentioned above, we also select other latest approaches for retraining: Pydiff \cite{ijcai}, NAFNet \cite{naf}, MIRNet \cite{mir} and RLP \cite{rlp}.

\subsubsection{Quantitative and Qualitative Evaluations.} 
Table \ref{table1} shows the average PSNR and SSIM values of the reported restoration methods on synthetic images from the LLR Dataset. The results demonstrate that our L$^{2}$RIRNet performs best among all the methods. Importantly, our approach shows a significant advantage compared to the eight cascade approaches. After retraining these state-of-the-art networks on the LLR Dataset, our L$^{2}$RIRNet still performs favorably against them. For instance, compared to DRSFormer \cite{drsformer}, there is a similar number of model parameters but increases 0.95 dB/0.0049 in PSNR and SSIM, respectively. As shown in Table \ref{table5}, the proposed L$^{2}$RIRNet achieves the lowest NIQE score and the highest CEIQ, indicating that our results are perceptually best.

Figs. \ref{zhenshi_1} and \ref{hecheng} provide a visual comparison of synthetic and real-world images from the LLR Dataset. Whether it is a synthetic or real image, the results of the cascade method are consistently unsatisfactory, often leaving behind rain pattern residue or overexposure. In real-world images, some of the latest methods struggle to remove rain streaks hidden in the darkness. In contrast, our method not only removes the rain streaks (first and fourth rows of Fig. \ref{zhenshi_1}) but also makes the image background clearer (second row of Fig. \ref{zhenshi_1}) and avoids blurring local scenes (third row of Fig. \ref{zhenshi_1}). For synthetic data, ours still achieves the best visual results.

\begin{table}
\begin{center} 
\fontsize{10}{10}\selectfont
\caption{\label{xiaorong_fdg}Effectiveness of the Fourier Detailed Guidance module.}
\setlength{\tabcolsep}{6mm}{
\begin{tabular}{cccccc}
\toprule
\myrowcolour%
Methods & PSNR (dB) & SSIM \\
\midrule
without DIG & 29.56& 0.8883\\
\myrowcolour%

without frequency & 30.12 & 0.8953\\
without spatial & 28.69 & 0.8842\\
\myrowcolour%

Ours & 30.53 & 0.9030\\
\midrule
\end{tabular}}%
\label{table7}
\end{center} 
\end{table}%

\begin{table}
\begin{center} 
\fontsize{10}{10}\selectfont
\caption{\label{xiaorong_ddl}Effectiveness of the dual degradation contrastive loss.}
\setlength{\tabcolsep}{5.3mm}{
\begin{tabular}{cccccc}
\toprule
\myrowcolour%
Methods & PSNR (dB) & SSIM \\
\midrule
Only Restoration-Net & 28.06 & 0.8762 \\
Patch-based (DASR) & 27.13 & 0.8542\\
\myrowcolour%
Image-based (Ours) & 30.53 & 0.9030 \\
\midrule
\end{tabular}}%
\label{table8}
\end{center} 
\end{table}%

\begin{table}
\begin{center} 
\fontsize{40}{40}\selectfont
\caption{\label{xiaorong_share}Effectiveness of forms of degradation representation learning network.} 
\resizebox{\linewidth}{!}{
\begin{tabular}{ccccc}
\toprule
\myrowcolour%

Methods & PSNR (dB) & SSIM & Params (M) & Runtime (Ms) \\
\midrule
DDR-unshared & 30.12 & 0.897 & 152.62 & 112.43\\
\myrowcolour%

DDR-shared (ours) & 30.53 & 0.903& 73.88 & 58.38\\
\midrule
\end{tabular}}%
\label{table6}
\end{center} 
\end{table}%

\subsection{Ablation Study}

\textbf{Effectiveness of key architecture.} 
Fig. \ref{xiaorong} shows five different variants of the model. V1 represents only Retoration-Net. V2 combines the Restoration-Net with the single degradation representation network (SDR-Net). V3 combines the Restoration-Net with the dual degradation representation network (DDR-Net). V4 represents our stage1 of L$^{2}$RIRNet, and V5 is our complete process with stage1 and stage2. 

By carefully comparing these variants, we gain important insights into the performance of our L$^{2}$RIRNet. Comparing V1 with V2, we find that adding a degradation representation guidance can improve restoration performance. Furthermore, comparing V2 with V3, the PSNR is increased by 0.58 dB. This result proves the vital role of our DDR-Net in extracting degradation information from both bright and dark regions using dual channels, resulting in significant improvement compared to single channels. Most importantly, when we compare V3 with V4, 0.84dB is increased in PSNR by adding the FDG module. This result highlights the value of our proposed model, as it demonstrates the importance of adding prior details and Fourier frequency domain information to the restoration process. The V5 increases by 0.57 dB in PSNR over the V4. These results demonstrate the effectiveness of each key part of our L$^{2}$RIRNet architecture and the importance of combining them to achieve satisfactory performance.

\textbf{Effectiveness of the Fourier Detailed Guidance Module.} In Table \ref{xiaorong_fdg}, we present our L$^{2}$RIRNet without detailed image generation (DIG), frequency, and spatial branch. All four versions are two-stage training methods. We can observe that the DIG module can enhance the performance from 29.56dB/0.8883 to 30.53dB/0.9030, which proves the validity of DIG. The FSFE, without frequency and spatial branches, is 30.12dB/0.8953 and 28.69dB/0.8842, separately.

\textbf{The Benefit of the Dual Degradation Contrastive Loss (DDLoss).} As shown in Table \ref{xiaorong_ddl}, our image-based strategy obtains 30.53dB/0.9030, achieving 2.47dB/0.0268 gains compared to only Restoration-Net. Meanwhile, the patch-based contrastive learning loss from DASR \cite{dasr} obtains 27.13dB/0.8542, even lower than only using Restoration-Net 28.06dB/0.8762. As shown in Fig. \ref{pdie_1}(a), the patch-based approach may lead to selecting blocks with different degradation representations. The incorrectly learned representations will result in an unsatisfactory representation learning network and thus affect the performance of the restoration.

\textbf{Effectiveness of Forms of Degradation Representation Network.} To validate the enhancement of our DDR-Net, we compare our dual-branch shared version with the dual-branch unshared version. Table \ref{xiaorong_share} shows that using a sharing DDR-Net can improve performance and reduce the number of parameters and runtime.

\section{CONCLUSION}
In this paper, we present a two-stage framework for low-light-rainy image restoration that performs favorably against existing solutions for both synthetic and real images. As a key design, we propose a DDR-Net that can extract distinct degradation representations for the degraded bright and dark regions, thereby accurately capturing the causes of image degradation. In addition, we introduce an FDG module based on detailed image and frequency spatial feature extraction that provides excellent prior guidance information for the restoration process. Our approach offers significant improvements over current methods, highlighting the effectiveness of our framework in addressing the challenges of low-light-rainy image restoration.

\section{Limitations and Future work}
L$^{2}$RIRNet efficiently solves the low-light-rainy image restoration task in synthetic and real-world scenes. Some diffusion-based models may obtain excellent results. Additionally, a degradation-robust semantic feature may provide better guidance as it focuses on the image content information, not incomprehensible degrading factors. So, in the future, a semantic-aware encoder or a large-scale pre-trained model that can provide stable features to guide the restoration process is worth exploring.
\bibliography{e}
\bibliographystyle{IEEEtran}

\end{document}